УДК 531.19

# РАСПРЕДЕЛЕНИЯ ПОТЕНЦИАЛА И КОНЦЕНТРАЦИИ НОСИТЕЛЕЙ ЗАРЯДА В ТВЕРДОТЕЛЬНОМ ЭЛЕКТРОЛИТЕ МЕЖДУ ПЛОСКИМИ ЭЛЕКТРОДАМИ

*Г. С. БОКУН*[1], *Д. ДИ КАПРИО*[2]

[1]*Белорусский государственный технологический университет,
ул. Свердлова, 13а, 220630, г. Минск, Беларусь*
[2]*Национальная высшая школа химии в Париже,
ул. Пьера и Марии Кюри, 11, 75005, г. Париж, Франция*

Статистически изучаются равновесные характеристики подсистемы подвижных зарядов одного знака с учетом наличия подсистемы неподвижных зарядов противоположного знака, создающих компенсирующий электрический фон. Распределение этих зарядов под воздействием внешнего поля не изменяется. Для представления свободной энергии подсистемы подвижных зарядов в виде функционала их плотности и вычисления ячеечных потенциалов средних сил методом условных распределений применено кумулянтное разложение по перенормированным майеровским функциям. Для учета эффектов экранирования использованы результаты метода коллективных переменных. Получена система интегральных уравнений для потенциалов средних сил с учетом эффектов близкого и дальнего действия, с помощью которой проведены расчеты в решеточном приближении. В выражении для бинарной функции распределения выделена корреляционная составляющая, что позволило рассчитать коррелированную и некоррелированную части электрического потенциала, используя уравнение Пуассона. Рассматривается случай достаточно малых электрических полей, допускающий линейное разложение химического потенциала по отклонению концентрации зарядов от однородного распределения. В окончательных расчетах корреляция между частицами учитывается в приближении первых соседей. В этом приближении распределение потенциала и концентрации заряда описывается линейным дифференциальным уравнением четвертого порядка. Выполнен анализ результатов его аналитического решения и последующих численных расчетов характеристик твердого электролита.

*Ключевые слова:* твердые электролиты; экранированный потенциал; близкодействие; функционал свободной энергии; электроемкость.

**Авторы:**
*Георгий Станиславович Бокун* – кандидат физико-математических наук; доцент кафедры механики и конструирования факультета химической технологии и техники.
*Дунг Ди Каприо* – доктор философии; научный сотрудник.

**Authors:**
*Heorhi S. Bokun*, PhD (physics and mathematics); associate professor at the department of mechanics and design, faculty of ChTiT.
*gbokun12@gmail.com*
*Dung Di Caprio*, PhD (philosophy); researcher.
*dung.di-caprio@chimie-paristech.fr*

# POTENTIAL AND CHARGE-CARRIER CONCENTRATION DISTRIBUTIONS IN SOLID ELECTROLYTE BETWEEN FLAT ELECTRODES

*H. S. BOKUN*[a], *D. DI CAPRIO*[b]

[a]*Belarusian State Technological University, 13a Sviardlova Street, Minsk 220630, Belarus*
[b]*National Graduate School of Chemistry, 11 Pierre and Marie Curie Street, Paris 75005, France*
*Corresponding author: H. S. Bokun (gbokun12@gmail.com)*

Statistically studied are the equilibrium characteristics of a subsystem of mobile charges of one sort, taking into account the subsystem of fixed charges of the opposite sign creating a compensating electric background. The distribution of these charges under the influence of the external field is invariable. To represent free energy of the subsystem of mobile charges in the form of a functional of their density and to calculate cell potentials of the mean forces by the method of conditional distributions, a cumulant expansion with respect to the renormalized Mayer functions is used. To take into account the screening effects, the results of the collective variables method are used. A system of integral equations for the potentials of mean forces is obtained that accounts for the effects of near- and long-range interactions. The calculations are made in the lattice approximation. The correlation component distinguished in the expression for the binary distribution function makes it possible to calculate the correlated and uncorrelated parts of the electric potential using the Poisson equation. In the case of sufficiently small electric fields, a linear expansion of the chemical potential in terms of deviation of the charge concentration from the homogeneous distribution is considered. In final calculations the correlation between particles is taken into account in the approximation of the first neighbors. In this approximation the potential and charge concentration distribution is described by a linear differential equation of the fourth order. The results of its analytical solution and subsequent numerical calculations for the characteristics of solid electrolyte are analyzed.

***Acknowledgements.*** The authors acknowledge financial support from the Belarusian Republican Foundation for Fundamental Research (grant No. Ф16К-061), the Ministry of Education of Belarus and the European Union's Horizon-2020 research and innovation programme under the Marie Skłodowska-Curie (grant agreement No. 73427 CONIN).

## Введение

Твердотельные керамические материалы являются ионными проводниками с преимущественной подвижностью зарядов одного знака. Они широко используются в электрохимических системах, в частности в высокотемпературных топливных элементах [1; 2], а также при разработке новых твердотельных электрохимических источников тока [3; 4] в целях повышения безопасности их эксплуатации за счет исключения из их конструкций жидких электролитов и ионных жидкостей. В связи с этим особую актуальность приобретает необходимость разработки надежных теоретических методов описания различных свойств таких материалов. Одним из проблемных факторов, усложняющих это описание, является наличие дальнодействующих межионных кулоновских взаимодействий, требующих разработки новых методов учета эффектов экранирования в ионных системах.

Предложенные ранее способы учета как короткодействующих, так и дальнодействующих взаимодействий в конденсированных средах [5] могут быть использованы для описания кристаллических ионных систем и токопроводящих керамик, основные особенности которых достаточно хорошо воспроизводятся решеточной теорией [6]. К таким способам можно при соответствующих аппроксимациях свести двухуровневый молекулярно-статистический метод описания неоднородных систем [7; 8], являющийся модификацией метода условных распределений [9; 10]. Основные характеристики систем, рассматриваемых в данной работе, рассчитываются с помощью кумулянтных (групповых) разложений по перенормированным майеровским функциям, когда в явном виде учитывается кристаллическая структура исследуемых материалов. Это позволяет описать эффекты корреляции в ионных системах с помощью функций распределения базисной системы, хотя последняя формально выглядит как некоррелированный физический объект.

## Большая статистическая сумма твердого электролита

Для вычисления большой статистической суммы изучаемой системы воспользуемся кумулянтным разложением по обобщенным майеровским функциям, формируемым на основе функций распределения базисной системы, гамильтониан которой представим модифицированными за счет учета вакансий в кристаллах одночастичными ячеечными потенциалами $\varphi_j(q_{n_i})$ метода условных распределений [7–10].

Переменная $q_{n_i}$ определяет положение частицы ($n_i = 1_i$) либо вакансии ($n_i = 0_i$) в $i$-й ячейке объемом $\omega_i$ ($i = 1, 2, \ldots, M$, где $M$ – общее число ячеек, на которые разделен весь объем $V$ изучаемой кристаллической системы; $V = \sum_{i=1}^{M} \omega_i$). Потенциал $\varphi_j(q_{n_i})$ имеет смысл потенциала локального поля, источник которого расположен в $j$-й ячейке решетки. Этот потенциал зависит не только от положения частицы в $i$-й ячейке, но и от параметров, характеризующих распределение частиц по объему $V$-системы. Гамильтониан $H_0$ базисной системы в формализме большого канонического ансамбля представим с помощью ячеечных потенциалов $\varphi_j(q_{n_i})$ в виде, который допускает (обеспечивает) факторизацию ее функции распределения для всех частиц:

$$H_0 = \sum_{i=1}^{M} \mu_i n_i + \sum_{i=1}^{M} \sum_{j(i)}^{Z} \varphi_j(q_{n_i}). \tag{1}$$

Суммирование по $j$ в (1) выполняется по всем узлам, окружающим избранный узел $i$ в пределах первых $Z$ координационных сфер.

Большая статистическая сумма $Z_V^{(0)}$ базисной системы имеет вид ($\beta = \frac{1}{kT}$):

$$Z_M^{(0)} = \sum_{n_1=0}^{1} \int_{\omega_1} dq_{n_1} \ldots \sum_{n_i=0}^{1} \int_{\omega_i} dq_{n_i} \ldots \sum_{n_M=0}^{1} \int_{\omega_M} dq_{n_M} \cdot \exp\left(-\beta\left(\sum_{l=1}^{M}\left(\mu_l n_l + U_{n_l}\right)\right)\right), \quad U_{n_l} = \sum_{j(l)}^{Z} \varphi_j(q_{n_l}). \tag{2}$$

Далее, рассматривается система с постоянной температурой, с неоднородностью, характеризуемой полем средних чисел заполнения ячеек системы $\rho_{n_i} = \langle n_i \rangle_0$, связанных с ячеечными химическими потенциалами $\mu_i$ соотношениями, которые вытекают из (2).

$$\frac{\rho_{1_i}}{\rho_{0_i}} = \exp(\beta \mu_i) \frac{Q_{1_i}}{Q_{0_i}}, \quad Q_{n_i} = \int_{\omega_i} \exp\left(-\beta \sum_{k(i)}^{z} \varphi_k(q_{n_i})\right) dq_{n_i}. \tag{3}$$

Поскольку в соответствии с представлением (1) функция распределения частиц базисной системы является факторизованной, это позволяет вычислять корреляторы, определяющие статистическую сумму реальной ионной системы. Ее гамильтониан, состоящий из парных вандерваальсовых короткодействующих потенциалов Ф и кулоновских дальнодействующих потенциалов $V$, записывается в следующем виде:

$$H_M = \frac{1}{2} \sum_{i=1}^{M} \sum_{j(=i)}^{Z} \Phi(q_{n_i}, q_{n_j}) + \frac{1}{2} \sum_{i=1}^{M} \sum_{j(=i)}^{Z} V(q_{n_i}, q_{n_j}) + \sum_{i=1}^{M} \mu_i n_i. \tag{4}$$

Выделяя из этого гамильтониана кулоновское взаимодействие, отклонение оставшейся части от гамильтониана базисной системы определим соотношениями:

$$\Delta H_M = \frac{1}{2} \sum_{i=1}^{M} \sum_{j(i)}^{Z} \Delta \varphi(q_{n_i}, q_{n_j}), \quad \Delta \varphi(q_{n_i}, q_{n_j}) = \Phi(q_{n_i}, q_{n_j}) - \varphi_j(q_{n_i}) - \varphi_i(q_{n_j}).$$

Статистическую сумму $Z_M$ системы с гамильтонианом (4) преобразуем с учетом выражения (1), в результате получим

$$Z_M = Z_M^{(0)} < e^{-\beta \Delta H_M} >_0. \tag{5}$$

Здесь обозначение $<\ldots>_0$ определяет усреднение, которое выполняется с помощью факторизованной функции распределения базисной системы, представленной произведением унарных функций $\hat{\rho}(q_{n_1})$. Это усреднение для произвольной функции $L$ имеет следующий вид:

$$<L>_0 = \sum_{n_1=0}^{1} \ldots \sum_{n_M=0}^{1} \int_{\omega_1} \hat{\rho}(q_{n_1}) dq_{n_1} \ldots \int_{\omega_M} \hat{\rho}(q_{n_M}) dq_{n_M} L \exp\left(-\frac{\beta}{2} \sum_{i=1}^{M} \sum_{j(=i)}^{Z} V(q_{n_i}, q_{n_j})\right),$$

где $\hat{\rho}(q_{n_i}) = \hat{F}_1(q_{n_i}) \rho_{n_i}$.

Для вычисления статистической суммы (5) воспользуемся разложением по базисному распределению, в результате приходим к разложению по обобщенным майеровским функциям $f(q_{n_i}, q_{n_j})$ конденсированного состояния в виде

$$f(q_{n_i}, q_{n_j}) = \exp(-\beta \Delta\varphi(q_{n_i}, q_{n_j})) - 1.$$

Оставляя два первых члена разложения, получим

$$Z_M = Z_M^{(0)}\left(1 + \frac{1}{2}\sum_{i=1}^{M}\sum_{j(i)}^{Z}\sum_{n_i=0}^{1}\sum_{n_j=0}^{1}\int_{\omega_i}dq_{n_i}\int_{\omega_j}dq_{n_j} \cdot \hat{\rho}(q_{n_i})\hat{\rho}(q_{n_j})g(q_{n_i}, q_{n_j})f(q_{n_i}, q_{n_j})\right), \quad (6)$$

где $g(q_{n_i}, q_{n_j})$ – бинарная функция распределения частиц в системе с дальнодействием, которое описывается одночастичными ячеечными потенциалами и входит в выражение для полной бинарной функции условных распределений:

$$\hat{\rho}(q_{n_i}, q_{n_j}) = \rho_{n_i}\rho_{n_j}\exp\{-\beta\Phi(q_{n_i}, q_{n_j})\} \cdot g(n_i, n_j)\hat{F}_1(q_{n_i})\hat{F}_1(q_{n_j})\exp\{-\beta(\varphi_i(q_{n_j}) + \varphi_j(q_{n_i}))\}.$$

Искомую статистическую сумму (6) преобразуем с учетом соотношения между полной унарной функцией $\hat{\rho}(q_{n_i})$ и нормированной на единицу унарной функцией $\hat{F}_1(q_{n_i})$ модифицированного метода условных распределений:

$$Z_M = Z_M^{(0)}\left(1 + \frac{1}{2}\sum_{i=1}^{M}\sum_{j(i)}^{Z}\sum_{n_i=0}^{1}\sum_{n_j=0}^{1}\rho_{n_i}\rho_{n_j}f_{n_i n_j}\right),$$

где

$$f_{n_i, n_j} = \int_{\omega_i}dq_{n_i}\int_{\omega_j}dq_{n_j}\,g(q_{n_i}, q_{n_j})f(q_{n_i}, q_{n_j})\hat{F}_1(q_{n_i})\hat{F}_1(q_{n_j}).$$

Уравнение для одночастичных потенциалов $\varphi_j(q_{n_i})$ получим из условия экстремальности для статсуммы с двумя оставленными членами в разложении (6). Это условие следует из независимости суммы всех членов разложения от выбора одночастичных потенциалов. Выполнив варьирование (6) по потенциалам $\varphi_j(q_{n_i})$, приходим к системе интегральных уравнений:

$$\exp(-\beta\varphi_k(q_{n_m})) =$$
$$= \frac{1}{z_k^0}\sum_{n_k=0}^{1}\exp(\beta\mu_k n_k)\int_{\omega_k}dq_{n_k}\,g(q_{n_m}, q_{n_k})\exp\left(-\beta\left(\Phi(q_{n_m}, q_{n_k}) + \sum_{s \neq m, k}^{Z}\varphi_s(q_{n_k})\right)\right), \quad (7)$$

где

$$z_k^0 = \sum_{n_k=0}^{1}\exp(\beta\mu_k n_k)\int_{\omega_k}dq_{n_k}\exp\left(-\beta\left(\sum_{s \neq m, k}^{Z}\varphi_s(q_{n_k})\right)\right).$$

Полученные интегральные уравнения (7) отличаются от аналогичных уравнений для молекулярных систем тем, что их ядра, помимо парного короткодействующего потенциала Ф, содержат бинарную функцию $g$ для системы частиц с кулоновским взаимодействием, приближенное выражение для которой получено в работе [5] и определяет корреляционную функцию $h(q_{n_i}, q_{n_j})$:

$$h(q_{n_i}, q_{n_j}) = g(q_{n_i}, q_{n_j}) - 1 = \exp(-\beta V_s(q_{n_i}, q_{n_j})) - 1, \quad (8)$$

где $V_s(q_{n_i}, q_{n_j})$ – уже не точечный кулоновский, а экранированный кулоновский потенциал. Таким образом, из (7) и (8) видно, что экранирование дальнодействия проявляется не только на кулоновском потенциале, но и на одночастичных ячеечных потенциалах базисной системы.

### Решеточное описание

Для перехода к решеточному описанию конденсированных систем нужно в интегральных уравнениях (7) и других формулах предыдущего раздела все функции координат заменить их значениями, определенными для случая, когда частицы удерживаются в центрах своих ячеек, т. е. в соответствующих узлах решетки. В связи с этим введем обозначения для значений, необходимых для дальнейших расчетов функций:

$$W_{n_i, n_j} = \exp(-\beta \Phi(n_i, n_j)), \tag{9}$$

$$G_{n_i, n_j} = \exp(-\beta V_s(n_i, n_j)), \tag{10}$$

$$U_{n_i, n_j} = W_{n_i, n_j} G_{n_i, n_j}, \tag{11}$$

$$f_{j, n_i} = \exp(-\beta \varphi_j(n_i)), \tag{12}$$

где $\Phi$, $V_s$, $\varphi$ – значения близкодействующего, экранированного и ячеечного потенциалов, вычисленные для соответствующих узлов решетки.

В результате интегральные уравнения (7) с учетом обозначений (3), (9)–(12) приобретают следующий вид:

$$f_{j, 1_i} = \frac{\rho_{1j} \cdot U_{1_i, 1_j}}{f_{i, 1_j}} + \frac{\rho_{0j}}{f_{i, 0_j}}, \quad f_{j, 0_i} = \frac{\rho_{1j}}{f_{i, 1_j}} + \frac{\rho_{0j}}{f_{i, 0_j}}.$$

Все термодинамические характеристики ионной твердотельной системы рассчитываются с помощью коррелятивных функций $\hat{\rho}(q_{n_i})$, $\hat{\rho}(q_{n_i}, q_{n_j})$, которые для решеточных систем выражаются через средние значения чисел заполнения и соответствующие величины из обозначений (9)–(12):

$$\hat{\rho}(n_i) = \rho_{n_i}, \quad \hat{\rho}(n_i, n_j) = \rho_{n_i} \rho_{n_j} \frac{U_{n_i, n_j} f_{i, 0_j} f_{j, 0_i}}{f_{i, n_j} f_{j, n_i} f_{i, 0_j} f_{j, 0_i}}. \tag{13}$$

В дальнейшем учтем, что проводимость в твердых электролитах обеспечивается подвижностью катионов, тогда как анионы остаются неподвижными и создают компенсирующее электрическое поле. При этом их распределение не изменяется под действием внешнего поля заряженного плоского конденсатора, создающего электрическое поле в направлении оси $x$ с напряженностью $E = \text{const}$ и потенциалом $\Psi^e = -Ex$.

Таким образом, при построении функционала свободной энергии твердого электролита необходимо учесть взаимодействие катионов только между собой и с внешним полем $\Psi^e(x)$ конденсатора. В рассматриваемом подходе статистическая сумма ионной системы представляется статистической суммой формально некоррелированного кристалла, но с переопределенными за счет корреляции одночастичными потенциалами. В связи с этим выражение для свободной энергии ионного кристалла в первом приближении составим в виде суммы внутренней энергии решеточной системы и комбинаторной части энтропии. Переходя к сокращенным обозначениям $c_i = \rho_{1_i}$ ($\rho_{0_i} = (1 - c_i)$), записываем с учетом (9)–(13) выражение для функционала свободной энергии $F$ подсистемы катионов во внешнем поле:

$$\beta F = \sum_{i=1}^{M} \left( c_i \ln(c_i) + (1 - c_i)_i \ln(1 - c_i) + r_b \sum_{j(i)} \left( \frac{c_j c_i}{r_{ij}} \right) + \beta \sum_{j \neq i} J_{ij} c_j c_i + \sum_{j \neq i} h_{ij} c_j c_i + \beta e c_i \Psi^e \right), \tag{14}$$

где $c_i = \rho_{1_i}$ – концентрация катионов; $(1 - c_i)$ – концентрация вакансий; $e$ – величина заряда иона; $r_b$ – радиус Бьеррума; $r_{ij}$ – расстояние между узлами $i$ и $j$; $J_{ij}$ – средняя энергия короткодействия между двумя частицами, фиксированными в узлах $i$ и $j$; $h_{ij}$ – параметр, определяющий коррелированную часть кулоновской энергии двух частиц, фиксированных в узлах $i$ и $j$, и выражаемый через функцию (10) по формуле

$$\tilde{h}_{ij} = \frac{r_b}{r_{ij}} (G_{1_i, 1_j} - 1), \quad r_b = \frac{\beta e^2}{4\pi \varepsilon \varepsilon_0}.$$

## Распределение электрического потенциала и заряда в твердом электролите в решеточном приближении

Рассмотрим случай слабого внешнего электрического поля, напряженность $E$ которого направлена вдоль оси $x$, действующего на кристаллический однородный образец твердого электролита. В связи с этим при варьировании свободной энергии (14) по полю с концентрацией $c_i$ изменением параметров $J_{ij}$ и $h_{ij}$ будем пренебрегать. В итоге выражение для химического потенциала $\mu_i = \dfrac{\delta F}{\delta c_i}$ подвижных катионов принимает следующий вид:

$$\beta \mu_i = \ln\left(\frac{c_i}{1-c_i}\right) + r_b \sum_{j \neq i}^{M}\left(\frac{c_j}{r_{ij}}\right) + \beta \sum_{j \neq i} J_{ij} c_j + \sum_{j \neq i} h_{ij} c_j + \beta e \Psi^e.$$

В случае отсутствия внешнего поля концентрация $c_i = c = \mathrm{const}$, тогда для соответствующего химического потенциала $\mu_i$ запишем, что

$$\beta \mu = \ln\left(\frac{c}{1-c}\right) + r_b \sum_{j \neq i}^{M}\left(\frac{c}{r_{ij}}\right) + \beta \sum_{j \neq i} J_{ij} c + \sum_{j \neq i} h_{ij} c. \qquad (15)$$

Далее, учтем, что под действием электрического поля конденсатора в объеме твердого электролита формируется антисимметричное одномерное распределение электрического потенциала $\Psi(x)$ и плотности заряда относительно срединной плоскости, параллельной плоским обкладкам конденсатора. Когда расстояние между обкладками существенно превосходит линейный размер областей неоднородности концентраций вблизи электродов, то в окрестности срединной плоскости будет наблюдаться область однородности с концентрацией, равной $c$, где химический потенциал определяется соотношением (15). В состоянии термодинамического равновесия химический потенциал $\mu_i$ одинаков во всех узлах системы и равен химическому потенциалу $\mu$. Приравняв $\mu_i$ и $\mu$, получим систему нелинейных уравнений, которая является дискретным аналогом интегрального уравнения для равновесного поля с концентрацией $c_i$ ($\delta c_i = c_i - c$):

$$\ln\left(\frac{c_i}{1-c_i}\right) - \ln\left(\frac{c}{1-c}\right) + r_b \sum_{j(i)}\left(\frac{\delta c_j}{r_{ij}}\right) + \beta \sum_{j \neq i} J_{ij} \delta c_j + \sum_{j \neq i} h_{ij} \delta c_j + \beta e \Psi^e = 0, \ i = 1, 2, \ldots. \qquad (16)$$

Поскольку $J_{ij}$ и $h_{ij}$ короткодействующие, то ограничимся учетом взаимодействия только с ближайшими соседями, положив $\beta J_{ij} = J$, $h_{ij} = h$.

Для некоррелированной части кулоновского взаимодействия введем потенциал

$$\hat{\Psi} = \beta e \Psi = r_b \sum \frac{\delta c_j}{r_{ij}},$$

для которого справедливо уравнение Пуассона с равновесным искомым полем с вариацией концентрации $\delta c_i$

$$\Delta \hat{\Psi} = -\frac{\beta e^2}{\varepsilon \varepsilon_0 H^3} \delta c, \qquad (17)$$

где $\Delta$ – оператор Лапласа; $H$ – параметр кристаллической решетки.

Переходя далее к линейным размерам в единицах параметра $H$ решетки, перепишем уравнение (17):

$$\Delta \hat{\Psi} = -u \delta c, \ u = \frac{\beta e^2}{\varepsilon \varepsilon_0 H}. \qquad (18)$$

Разложив первое логарифмическое слагаемое уравнения (16) по вариации концентрации $\delta c_i$, перепишем уравнение (16) в линейном приближении:

$$\frac{\delta c_i}{c(1-c)} + \hat{\Psi} + J_\Sigma \sum_{j \neq i} \delta c_j + \hat{\Psi}^e = 0, \ J_\Sigma = J + h.$$

Из суммы, содержащейся в последнем уравнении, выделим вторую численную разность $\Delta \delta c_i = \delta c_{i+1} - 2\delta c_i + \delta c_{i+1}$ и преобразуем его в дифференциальное уравнение ($z$ – число ближайших соседей для узлов используемой кристаллической решетки):

$$\gamma_T \delta c + \hat{\Psi} + J_\Sigma \Delta \delta c + \hat{\Psi}^e = 0, \quad \gamma_T = \frac{1}{c(1-c)} + z J_\Sigma. \tag{19}$$

С учетом уравнения Пуассона (18) уравнение (19) превращается в неоднородное дифференциальное уравнение четвертого порядка для потенциала $\hat{\Psi}$:

$$\Delta^2 \hat{\Psi} + a\Delta\hat{\Psi} - b\hat{\Psi} = -bEx, \quad a = \frac{\gamma_T}{J_\Sigma}, \quad b = \frac{u}{J_\Sigma}. \tag{20}$$

В связи с отмеченной выше антисимметрией для искомых полей решение уравнения (20) для потенциала $\hat{\Psi}$ в системе координат с началом в срединной плоскости межэлектродного пространства имеет вид

$$\hat{\Psi} = C_1\left(e^{k_1 x} - e^{-k_1 x}\right) + C_2 \sin(k_2 x) + Ex. \tag{21}$$

Входящие в уравнение (21) корни $k_1$, $k_2$ характеристического уравнения соответственно равны

$$k_1 = \sqrt{\frac{a}{2}\left(-1 + \sqrt{1 + \frac{4b}{a^2}}\right)}, \quad k_2 = \sqrt{\frac{a}{2}\left(1 + \sqrt{1 + \frac{4b}{a^2}}\right)}. \tag{22}$$

Если ограничиваться только вкладом от кулоновского взаимодействия без учета корреляций между ионами, т. е. положить в развиваемом подходе $J_\Sigma$ равным нулю, то выражение для первого корня приобретает вид

$$k_1 = \sqrt{c(1-c)u} = k_D.$$

Значение $k_D$ в решеточном приближении совпадает с полученным ранее выражением [11; 12] для обратного радиуса Дебая, входящего в экранированный кулоновский потенциал, определяющий согласно (8) корреляционную функцию $h$. Ее значение для ближайших соседей записывается в форме

$$h = -u \exp(-u k_D).$$

Для удобства выполнения дальнейших численных расчетов введем параметр $S_h$, определяющий отношение параметров короткодействия ($J$) и дальнодействия ($u$) для катионов в ближайших узлах решетки, т. е. $S_h = \dfrac{J}{u}$.

С учетом введенных обозначений запишем выражения для параметров $\gamma_T$ и $J_\Sigma$:

$$\gamma_T = \frac{1}{c(1-c)} + zu\left(S_h - \exp(-u k_D)\right), \quad J_\Sigma = u\left(S_h - \exp(-u k_D)\right).$$

Значение постоянной $C_2$ в уравнении (21) определяем из условия равенства нулю суммарной напряженности поля в начале координат, определяемой по формуле

$$E_\Sigma = -\frac{d\left(\hat{\Psi} + \hat{\Psi}^e\right)}{dx}.$$

В результате найдем $C_2$ и получим

$$\hat{\Psi} = C_1\left(\left(e^{k_1 x} - e^{-k_1 x}\right) - \frac{2k_1}{k_2}\sin(k_2 x)\right) + Ex. \tag{23}$$

Значение постоянной $C_1$ определяем из условия равенства в точке с координатой $x = L$ ($L$ – расстояние между обкладками конденсатора) заряда, распределенного в диффузионной области, и заряда на обкладке конденсатора.

В результате получим, что

$$\hat{\Psi}'(L) - \hat{\Psi}'(0) = E. \tag{24}$$

Определив постоянную интегрирования $C_1$ из уравнения (24), получаем окончательное выражение для суммарного потенциала $\hat{\Psi}_\Sigma$ в объеме твердого электролита:

$$\hat{\Psi}_\Sigma = \frac{E\left(\left(e^{k_1 x} - e^{-k_1 x}\right) - \dfrac{2k_1}{k_2}\sin(k_2 x)\right)}{k_1\left(\left(e^{k_1 L} + e^{-k_1 L}\right) - 2\cos(k_2 L)\right)}. \quad (25)$$

С помощью уравнения (25) находим электроемкость $C$ конденсатора с твердым электролитом в межэлектродном пространстве:

$$C = \frac{k_1\left(\left(e^{k_1 L} + e^{-k_1 L}\right) - 2\cos(k_2 L)\right)}{\left(e^{k_1 L} - e^{-k_1 L}\right) - \dfrac{2k_1}{k_2}\sin(k_2 L)},$$

а также распределение отклонения концентрации заряда от равновесного значения

$$\delta c(x) = -\frac{\Psi''}{u}. \quad (26)$$

Все расчеты выполнены при фиксированных значениях параметра $S_h$ ($S_h = 2$) и концентрации $c$ ($c = 0{,}5$). В этом случае все характеристики твердого электролита в приэлектродной области зависят от переменной $u$, которая для конкретного образца с диэлектрической проницаемостью $\varepsilon$ обратно пропорциональна температуре ($\beta = \dfrac{1}{kT}$). Результаты численных расчетов корней $k_1$, $k_2$ характеристического уравнения по формулам (22) представлены на рис. 1.

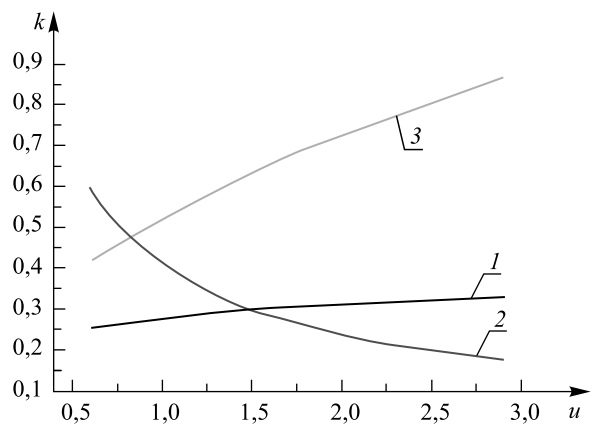

*Рис. 1.* Зависимости волновых чисел от энергии кулоновского взаимодействия
ближайших соседей в единицах абсолютной температуры:
*1* – $k_1$; *2* – $k^* = k_2 - 2{,}5$; *3* – $k_D$
*Fig. 1.* Wave numbers versus the Coulomb energy
of the nearest neighbors in units of absolute temperature:
*1* – $k_1$; *2* – $k^* = k_2 - 2.5$; *3* – $k_D$

Значения корня $k_1$ (см. рис. 1, кривая *1*) очень медленно увеличиваются при возрастании переменной $u$, т. е. при уменьшении температуры, тогда как аналогичный параметр $k_D$ в теории Дебая (см. рис. 1, кривая *3*) возрастает намного быстрее и имеет значения, в 2–3 раза превышающие значения параметра $k_1$. Уменьшение параметра $k_1$ по сравнению с параметром $k_D$ является следствием учета корреляции в распределении заряда в приэлектродной области. Второй корень $k_2$ (см. рис. 1, кривая *2*), описывающий волновой характер изменения потенциала $\hat{\Psi}$, имеет смысл волнового числа для соответствующего частного решения с постоянной интегрирования $C_2$ в уравнении (21). Завышенные значения $k_2$ при малых значениях $u$ (высоких температурах) являются следствием учета только линейного члена в разложении логарифма, которое выполнено при переходе от нелинейного уравнения (16) к линейному.

Действительно, из выражения (23) видно, что гармоническая составляющая, вызывающая колебания потенциала, показанного на рис. 2 (кривая *1*), содержит множитель, обратно пропорциональный $k_2$, что приводит к уменьшению вклада соответствующей составляющей потенциала при возрастании частоты колебаний. Но, как видно из (26), вариация концентрации (см. рис. 2, кривая *2*) оказывается прямо пропорциональной $k_2$ и соответственно имеет гораздо более выраженную амплитуду колебаний.

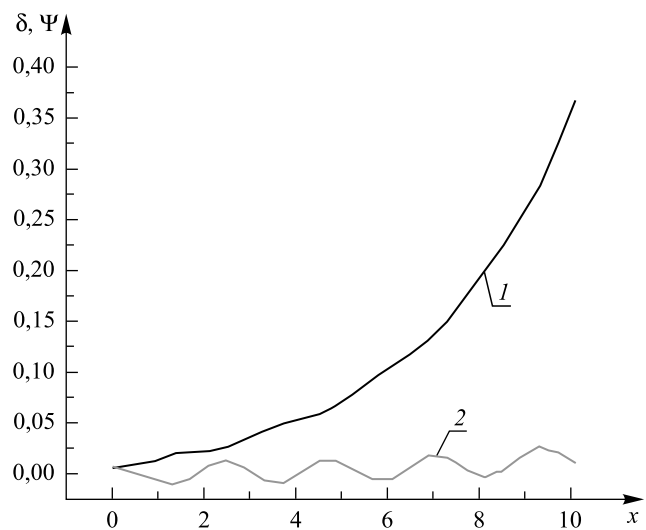

*Рис. 2.* Зависимость изменения электрического потенциала (*1*)
и концентрации носителей заряда (*2*) от расстояния до поверхности электрода

*Fig. 2.* Variations in the (*2*) concentration of charge carriers
and electric potential (*1*) versus the distance to the electrode surface

В целом в области конденсированного состояния при значениях *u*, превышающих единицу, вклад гармонической составляющей приводит к возникновению периодических колебаний концентрации, показанных на рис. 2 (кривая *2*). Эти закономерности, касающиеся структуризации распределения заряда и потенциала в пространстве ионного твердотельного проводника с преимущественной подвижностью зарядов одного знака, качественно соответствуют результатам, полученным для случая мезоскопических систем в работе [13].

## Заключение

На основании разложения большой статистической суммы по корреляциям, с использованием в качестве базисного распределения, характерного для идеального кристалла, получена свободная энергия в виде функционала плотности и ячеечных потенциалов средних сил. Для описания эффектов экранирования в твердом теле применен метод коллективных переменных. В рамках решеточного приближения получено выражение для химического потенциала неоднородной твердотельной электрохимической системы.

В результате установлено распределение заряда и электрического потенциала в твердоэлектролитной системе с преимущественной подвижностью одноименных ионов. Показано, что в таких системах на известное экспоненциальное затухание потенциала накладывается гармоническая составляющая, отражающая возникновение волнообразного распределения плотности заряда в приэлектродном пространстве. Установлена зависимость параметров, характеризующих затухание и периодичность, от различных условий. Изучена их зависимость от температуры и от отношения энергий близко- и дальнодействия для ближайших соседей.